\documentclass[sigconf]{acmart}

\AtBeginDocument{%
  }
\copyrightyear{2026}
\acmYear{2026}
\setcopyright{cc}
\setcctype{by}
\acmConference[CIKM '26]{Proceedings of the 35th ACM International Conference on Information and Knowledge Management}{November 07--11, 2026}{Rome, Italy}
\acmBooktitle{Proceedings of the 35th ACM International Conference on Information and Knowledge Management (CIKM '26), November 07--11, 2026, Rome, Italy}
\acmDOI{10.1145/3799682.3840706}
                
\acmISBN{979-8-4007-2539-5/2026/11}

\usepackage[utf8]{inputenc}
\usepackage[T1]{fontenc}
\usepackage{booktabs}
\usepackage{graphicx}
\usepackage{multirow}
\usepackage{xspace}
\usepackage{algorithm}

\usepackage{enumitem}
\usepackage{nicefrac}
\usepackage{microtype}
\usepackage{xcolor}
\usepackage{float}
\usepackage{amsmath}

\usepackage{amssymb}
\usepackage{amsthm}

\newcommand{\sndcg}{\text{sNDCG@5}}

\title{Risk-Aware Reranking for Agentic Tool Retrieval}

\author{Qinfei Li}
\affiliation{%
  \institution{University of Science and Technology of China}
  \city{Hefei}
  \country{China}
}
\email{lqfff1984@gmail.com}

\author{Xiaoxuan Dong}
\affiliation{%
  \institution{University of Electronic Science and Technology of China}
  \city{Chengdu}
  \country{China}
}
\email{202522010524@std.uestc.edu.cn}

\author{Jin Zhang}
\affiliation{%
  \institution{Lanzhou University}
  \city{Lanzhou}
  \country{China}
}
\email{mjzj35723@gmail.com}

\author{Dexu Yu}
\affiliation{%
  \institution{Fenz.AI}
  \city{Palo Alto}\country{United States}
}
\email{yu.dex@northeastern.edu}

\author{Wenhao Deng}
\affiliation{%
  \institution{University of Glasgow}
  \city{Glasgow}
  \country{United Kingdom}
}
\email{w.deng.1@research.gla.ac.uk}

\author{Junchen Fu}
\affiliation{%
  \institution{University of Glasgow}
  \city{Glasgow}
  \country{United Kingdom}
}
\email{j.fu.3@research.gla.ac.uk}

\author{Youhua Li}
\affiliation{%
  \institution{City University of Hong Kong}
  \city{Hong Kong}
  \country{China}
}
\email{youhuali2-c@my.cityu.edu.hk}

\author{Hanwen Du}
\affiliation{%
  \institution{The Ohio State University}
  \city{Columbus}
  \country{United States}
}
\email{du.1128@osu.edu}

\author{Chunxiao Li}
\authornote{Corresponding author.}
\affiliation{%
  \institution{University of Science and Technology of China}
  \city{Hefei}
  \country{China}
}
\email{chunxiao.li@ustc.edu.cn}

\renewcommand{\shortauthors}{Qinfei Li et al.}

\begin{document}

\begin{abstract}

Tool retrieval determines which external tools are exposed to an LLM agent
for a user query or task, making retrieval a critical pre-execution safety
boundary. Unlike document retrieval, tool retrieval exposes executable actions:
a tool that is useful for one task may be unnecessary or risky for another.
However, existing tool-retrieval methods primarily optimize semantic relevance,
and safety evaluations often focus on failures after tool execution rather than
risks introduced during retrieval.
We study risk-aware tool retrieval, where the goal is to retrieve useful tools
while reducing exposure to higher-risk tools. We propose a lightweight
reranking framework on top of a frozen first-stage retriever. The framework models query-conditioned relevance and tool-level exposure
risk separately, combines them through an explicit parameter controlling
the tradeoff between safety and utility, smooths scores over a ToolGraph,
and optionally applies rule-based safety constraints.
To support retrieval-time safety evaluation, we annotate 6,108 tools across
UltraTool and Seal-Tools with five ordinal risk levels and define metrics that
measure risky-tool exposure in the top-$k$ results.
Experiments on UltraTool and Seal-Tools show that our approach improves the
relevance--safety tradeoff over relevance-only retrievers and reranking
baselines, with the rule-filtered variant providing a conservative operating
point for safety-critical deployments. These findings indicate that retrieval-stage filtering can reduce the
candidate action space exposed to agents before execution, complementing
downstream tool-use safeguards. The code and supplementary materials are available at:
\href{https://github.com/qli447/risk-aware-tool-retrieval-release}
{\textcolor{blue}{\nolinkurl{https://github.com/qli447/risk-aware-tool-retrieval-release}}}.

\end{abstract}



\maketitle

\section{Introduction}
\label{sec:intro}

Large language model (LLM) agents increasingly rely on external tools
such as APIs, code executors, and file managers to accomplish complex
real-world tasks~\citep{qin2024toolllm,patil2024gorilla,shen2023hugginggpt,han2025video,wu2026gift}.
Given a user query or task, a tool retriever selects the top-$k$ candidate
tools from a library for the agent to invoke~\citep{shi2025retrieval}. In this sense, tool retrieval determines the agent's task-specific tool
exposure before it acts. Recent
task-specific retrieval models have substantially improved retrieval
accuracy; for example, ToolRet~\citep{shi2025retrieval} fine-tunes a dense
encoder on 43k tools and outperforms general-purpose retrievers by a large
margin. However, unlike document retrieval, where results are retrieved for
the model to read, tool retrieval exposes executable actions to the agent,
making retrieval errors potentially irreversible~\citep{yu2025safety}.
Tool retrieval should not rely solely on semantic relevance to the user query;
it should also account for the operational risks of retrieved tools. Once risky
tools are placed in the top-$k$, they may lead to consequences such as
unauthorized data deletion or credential exposure
\citep{yuan2024r,zhang2024agent}.
\begin{figure}[t]
  \centering
  \includegraphics[
    width=0.95\linewidth,
    trim = 40 0 80 5,
    clip
  ]{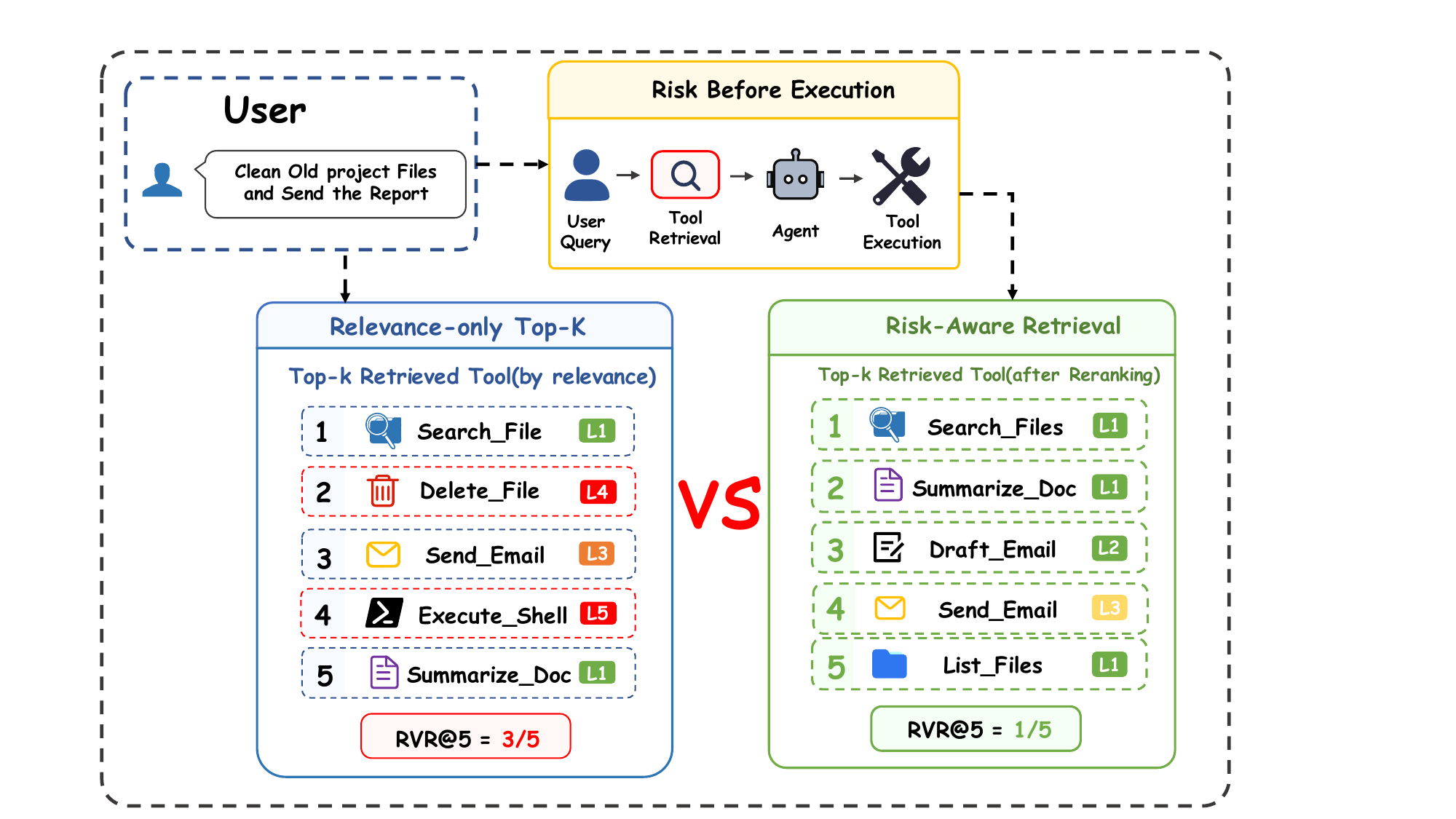}
  
  \caption{Tool retrieval as a pre-execution safety boundary. Relevance-only
top-$k$ retrieval exposes risky executable tools, while risk-aware reranking
reduces risky-tool exposure.}
  \Description{Three-panel figure showing relevance-only retrieval exposing
  high-risk tools before execution, while risk-aware reranking reduces the
  number of high-risk tools in the top-five list.}
  \label{fig:motivation}
\end{figure}
Figure~\ref{fig:motivation} illustrates this exposure gap: two top-$k$
lists can both contain useful tools, yet expose very different numbers of
higher-risk executable actions to the agent. The current tool retrieval pipeline has the following limitations.

\textit{i) } Existing retrievers and rerankers primarily optimize relevance,
not safe tool exposure. From BM25~\citep{beaulieu1997okapi} and dense
retrievers~\citep{xiao2024c,shi2025retrieval} to recent reranking
modules~\citep{yu2024rankrag,zheng2024toolrerank}, the goal is usually to
rank tools that match the query. However, executable tools are not passive
documents: a semantically relevant tool may still enable high-impact actions
such as credential access, file deletion, or code execution.
\textit{ii)} Standard retrieval metrics do not capture risky-tool exposure.
NDCG and MRR measure whether relevant tools appear near the top, but they do
not distinguish low-risk tools from high-risk ones. As a result, two rankings
can obtain similar relevance scores while exposing very different levels of
operational risk in the top-$k$.
\textit{iii)} Existing safety evaluations largely check safety after tool
selection or execution, rather than at the retrieval stage where the candidate
set is first formed~\citep{yuan2024r,zhang2024agent}. If the
retrieved candidates already contain many risky tools, downstream safety
checking has a much harder problem; in the extreme case, most exposed options
may be unsafe. Our experiments in Sections~\ref{sec:main_results}--\ref{sec:robustness}
show that relevance-only retrieval can expose substantially more higher-risk
tools than risk-aware alternatives.

Based on the above limitations, our goal is to make tool retrieval
risk-aware without modifying the upstream retriever.

\textit{First,} we introduce a dual-head reranker to move beyond relevance-only
tool ranking. A frozen ToolRet-BGE encoder feeds a relevance head
$f_{\mathrm{rel}}(q,t)$ and a tool-risk head $f_{\mathrm{risk}}(t)$, which
are combined as
$s(q,t)=f_{\mathrm{rel}}(q,t)-\lambda f_{\mathrm{risk}}(t)$.
This design exposes a controllable safety--utility tradeoff while keeping
the upstream retriever fixed.

\textit{Second,} we make retrieval-time safety measurable. We annotate 6,108 tools
with five ordinal risk levels and define RVR, SRR, and sNDCG to quantify
risky-tool exposure and safety-adjusted ranking quality in the top-$k$ list,
complementing relevance metrics such as NDCG and MRR.

\textit{Third,} we add pre-execution controls at the retrieval stage. A four-type
ToolGraph smooths scores among related tools, and an optional rule filter
enforces risk caps, permission constraints, and redundancy removal for
deployments that require stricter exposure control. The context-aware and
stress-test analyses in Sections~\ref{sec:robustness} and~\ref{sec:case_study}
further show why retrieval-stage control is needed.

Our contributions are summarized as follows:
\begin{itemize}[leftmargin=1.5em,itemsep=2pt,topsep=3pt]
  \item We formulate agent tool retrieval as a pre-execution
        relevance--safety optimization problem, where the goal is to retrieve
        useful tools while reducing exposure to higher-risk executable actions.

  \item We propose a lightweight risk-aware reranking framework on top of a
        frozen first-stage retriever. The framework combines dual-head
        relevance--risk scoring, graph-based score smoothing, and an optional
        rule filter for stricter deployment-time safety control.

  \item We annotate 6,108 tools across UltraTool and Seal-Tools with
        five-level operational risk labels, and define retrieval-time safety
        metrics including RVR, SRR, and sNDCG to measure risky-tool exposure
        and safety-adjusted ranking quality in the top-$k$ list.

  \item We evaluate the framework on UltraTool and Seal-Tools against 8
        retrieval and reranking baselines. Across main, robustness, and stress-test settings, the proposed reranker
        reduces risky-tool exposure while preserving competitive retrieval quality.
\end{itemize}

\section{Related Work}
\label{sec:related}

\textbf{Tool Retrieval.}
Tool retrieval and selection have been studied through tool-augmented LLM systems
and API retrieval~\citep{qin2024toolllm,patil2024gorilla,shen2023hugginggpt},
dense retrievers~\citep{shi2025retrieval,yuan2024craft}, generative tool
selection~\citep{wang2025toolgen}, query rewriting and reasoning
~\citep{chen2024re,sengupta2026tooldreamer}, and reranking methods
~\citep{zheng2024toolrerank}. These methods improve retrieval accuracy by
better matching user queries to useful tools. For example,
ToolRet~\citep{shi2025retrieval} fine-tunes a dense encoder on large-scale tool
data, while ToolRerank~\citep{zheng2024toolrerank} improves ranking with
cross-encoder scoring. However, existing methods primarily optimize relevance
or task success, without explicitly modeling the operational risk of exposing
retrieved tools. This gap is especially important for agents, where retrieved
items are executable actions rather than passive documents.

\textbf{Agent Safety Evaluation.} 
Many benchmarks evaluate the safety of tool-augmented LLM agents.
R-Judge~\citep{yuan2024r} studies safety-risk awareness in multi-turn
agent interactions, Agent-SafetyBench~\citep{zhang2024agent} evaluates
unsafe tool-use recognition, and ToolEmu~\citep{ruan2024identifying} tests
tool-call consequences in simulated environments. Other benchmarks, such as
AgentHarm~\citep{andriushchenko2025agentharm} and
SafeArena~\citep{tur2025safearena}, evaluate agents under harmful or
adversarial settings. These works provide valuable safety taxonomies, but they
primarily evaluate downstream agent decisions or execution outcomes rather than
risk exposure in the retrieved top-$k$ tool set. In addition, \citet{yu2025safety} show that retrieval
augmentation itself can degrade agent safety, indicating that risks are not
controlled during the retrieval process.

\textbf{Retrieval Reranking.}
Reranking is a common refinement step in retrieval pipelines. Existing
rerankers include cross-encoders~\citep{wang2020minilm,xiao2024c},
sequence-to-sequence models~\citep{nogueira2020document}, and
instruction-tuned LLM rerankers~\citep{yu2024rankrag}. These rerankers
typically optimize relevance, leaving safety as an external consideration
rather than an explicit ranking objective. In contrast, our work studies
retrieval-time relevance--safety optimization for executable tool candidates.

\textbf{Constrained and Diversified Ranking.}
Our formulation is also related to ranking under constraints and tradeoffs.
Diversified reranking methods such as maximal marginal relevance balance
query relevance with redundancy reduction~\citep{carbonell1998use}, while
fair top-$k$ ranking and exposure-aware ranking study how to optimize ranking
utility subject to distributional or exposure constraints~\citep{zehlike2017fa}. These works show that ranking objectives often need to
balance relevance with additional deployment constraints. Our setting differs
in that the constraint concerns operational risk of executable tools rather
than document novelty or group exposure.
\section{Preliminaries}
\label{sec:prelim}

Let $\mathcal{T}=\{t_1,\dots,t_N\}$ denote a tool corpus. Each tool
$t_i$ has a text description $d_i$ and an operational risk label
$r_i\in\{1,2,3,4,5\}$, where larger values indicate higher potential
impact if the tool is exposed to an agent. Given a user query
$q$, a first-stage retriever returns a candidate set
$\mathcal{C}_q\subset\mathcal{T}$, typically the top-100 tools by relevance.
A reranker then assigns a score $s(q,t)$ to each tool in the scored set---either $\mathcal{C}_q$ or the full corpus $\mathcal{T}$---and outputs a ranked top-$k$ list $\sigma_q^{(k)}$.

Unlike document retrieval, the retrieved items here are executable tools.
Therefore, the top-$k$ list is not only a relevance result but also the
candidate action space exposed to the downstream LLM agent. Our goal is to
learn a reranking function that maintains retrieval relevance while reducing
the exposure of higher-risk tools in $\sigma_q^{(k)}$. This setting focuses
on \emph{pre-execution} safety: no tool is executed during retrieval, but the
retrieved set constrains what the agent can choose to invoke next.

Formally, for each query $q$ with ground-truth relevant tools
$\mathcal{R}_q$, the reranker should rank relevant tools near the top while
discouraging unnecessary exposure of tools with high operational risk. We do
not impose a hard constraint during training; instead, the method exposes an
inference-time tradeoff parameter that allows practitioners to select different
relevance--safety operating points.

\section{Method}
\label{sec:method}

\subsection{Overview and Deployment Setting}
\label{sec:method_overview}

\begin{figure*}[t]
  \centering
  \includegraphics[width=\textwidth]{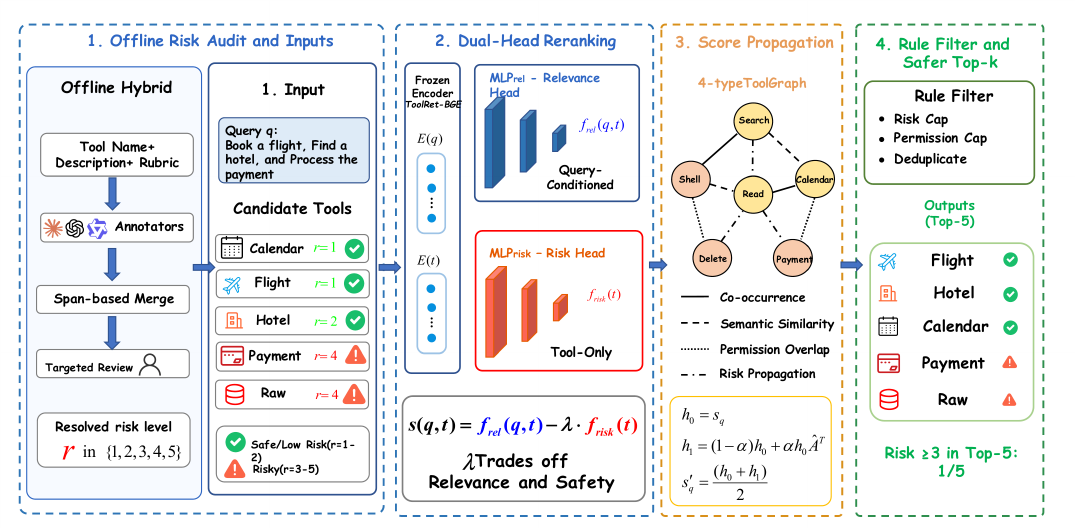}
  \caption{
  Overview of the risk-aware reranking framework.
  \textbf{Step~1:} An offline risk audit assigns ordinal risk labels to tools.
  \textbf{Step~2:} A frozen encoder feeds a query-conditioned relevance head
  and a tool-level risk head; their outputs are combined by an inference-time
  tradeoff parameter.
  \textbf{Step~3:} A ToolGraph smooths scores among related candidate tools.
  \textbf{Step~4:} When audited metadata is available and stricter constraints
  are required, an optional rule filter produces the final top-$k$ list.
  }
  \Description{A four-stage pipeline consisting of offline risk annotation,
  dual-head reranking, ToolGraph score smoothing, and an optional rule filter.}
  \label{fig:architecture}
\end{figure*}

Figure~\ref{fig:architecture} gives an overview of the framework. Given a
query $q$, a first-stage retriever returns a candidate set
$\mathcal{C}_q \subset \mathcal{T}$, typically the top-100 tools. Our method
can score either this candidate set or the complete tool pool, leaving the upstream retriever unchanged.

The framework has two operating modes. The \emph{core reranker} uses a
learned dual-head scoring model and graph-based score smoothing. At inference
time, this core mode uses the predicted risk score $f_{\mathrm{risk}}(t)$,
not the annotated risk label itself. The \emph{rule-filtered} mode adds a
deployment-time filtering layer on top of the core reranker. This mode assumes
an audited tool registry in which risk metadata and permission categories are
available before deployment, and is intended for settings that require stricter exposure constraints.

\subsection{Offline Risk Audit and Label Use}
\label{sec:annotation}

\textbf{Risk rubric.}
Each tool is assigned an ordinal operational risk level
$r_i\in\{1,\dots,5\}$ according to the potential operational consequence of exposing the tool without considering particular calls or runtime environments:

\begin{itemize}[leftmargin=1.5em,itemsep=1pt,topsep=2pt]
  \item \textbf{L1 Safe:} Read-only tools with no side effects or sensitive access.
  \item \textbf{L2 Low:} Minor reversible actions or non-sensitive personal-data access.
  \item \textbf{L3 Medium:} Sensitive data access or persistent writes.
  \item \textbf{L4 High:} Irreversible actions, security controls, or system-level permissions.
  \item \textbf{L5 Critical:} Large-scale harm, system intrusion, or severe privacy loss.
\end{itemize}

\textbf{Annotation procedure.}
Each tool is first labeled from its name and description by three LLM
annotators: Claude Code, Codex, and Qwen. Let the three votes be $v_1,v_2,v_3$, and define
$\mathrm{span}=\max(v_1,v_2,v_3)-\min(v_1,v_2,v_3)$.
We retain the label if $\mathrm{span}=0$, use the median if
$\mathrm{span}=1$, and have the authors review cases with
$\mathrm{span}\geq 2$.
A human researcher with expertise in LLM agents and tool use,
blinded to the released labels, independently labels a uniform random
sample of 150 tools from the pooled tool set.
Agreement with the released labels is 60.0\% exact and 77.3\% within
one level (quadratically weighted Cohen's $\kappa=0.362$).
Disagreements of two or more levels mainly concern underspecified access
to personal, financial, or health data, for which the human auditor uses
a more conservative risk level.

\textbf{Use of labels.}
The resolved labels have four roles. First, they provide offline supervision
for the risk head. Second, they define risk-related metadata used to construct
ToolGraph edges in audited tool libraries. Third, they support the optional
rule filter when deployment-time metadata is available. Fourth, they are used
for evaluation metrics such as RVR@5 and SRR@5. The core dual-head reranker
does not directly look up $r_i$ at inference time; it uses the predicted score
$f_{\mathrm{risk}}(t)$. The rule-filtered variant should therefore be read as
an audited-library deployment setting.

\begin{table}[t]
\caption{
Risk-label agreement and disagreement resolution. Span denotes the maximum
pairwise difference among the three LLM annotator scores. Agreement is measured by Krippendorff's $\alpha$.
}
\label{tab:annotation_agreement}
\centering
\resizebox{\columnwidth}{!}{%
\begin{tabular}{lccccc}
\toprule
Dataset & $\alpha_{\mathrm{K,all}}$ & $\alpha_{\mathrm{K,span}\leq 1}$
& span$=0$ & span$=1$ & span$\geq2$ \\
\midrule
UltraTool
& 0.523 & 0.707 & 928 (45.7\%) & 767 (37.7\%) & 337 (16.6\%) \\
Seal-Tools
& 0.856 & 0.901 & 3440 (84.4\%) & 196 (4.8\%) & 440 (10.8\%) \\
\bottomrule
\end{tabular}}
\end{table}

Table~\ref{tab:annotation_agreement} summarizes the annotation reliability.
UltraTool has lower agreement because its tools are more heterogeneous and
often described at a higher level of abstraction. Seal-Tools has more
standardized API-style descriptions and higher exact agreement. In both
datasets, most tools fall into either exact agreement or adjacent-level
disagreement; broad cross-level disagreements are manually reviewed.

\subsection{Dual-Head Risk-Aware Reranking}
\label{sec:dualhead}

\textbf{Scoring heads.}
A frozen ToolRet-BGE encoder maps the query and each candidate tool to
embeddings $\mathbf{e}_q$ and $\mathbf{e}_t$. We train two lightweight heads
on top of these frozen representations:
\begin{align}
  f_{\mathrm{rel}}(q,t)
  &= \mathrm{MLP}_{\mathrm{rel}}\bigl([\mathbf{e}_q;\mathbf{e}_t]\bigr),
  \label{eq:frel} \\
  f_{\mathrm{risk}}(t)
  &= \mathrm{MLP}_{\mathrm{risk}}\bigl(\mathbf{e}_t\bigr),
  \label{eq:frisk}
\end{align}
where $[\,\cdot\,;\,\cdot\,]$ denotes concatenation. Both heads use sigmoid
outputs, so their scores lie in the same numeric range. The relevance head is
query-conditioned, since the same tool may be essential for one query and
irrelevant for another. The risk head receives only the tool embedding and
estimates exposure risk at the tool level. It does not determine whether
a particular call is safe given the query, arguments, or environment. Thus, the combined score depends on the query, while the risk estimate does not. This estimate does not model argument-level or environment-dependent execution hazards; those effects are outside the information available to a retrieval-time reranker and remain the role of
downstream execution safeguards.

\textbf{Training objective.}
The two heads are trained jointly:
\begin{equation}
  \mathcal{L}
  = \mathcal{L}_{\mathrm{rel}}
  + \mu\,\mathcal{L}_{\mathrm{risk}},
  \label{eq:loss}
\end{equation}
where $\mu$ balances relevance learning and risk prediction. The relevance
loss is a pairwise margin loss over triplets $(q,t^+,t^-)$:
\begin{equation}
  \mathcal{L}_{\mathrm{rel}}
  =
  \sum_{(q,t^+,t^-)}
  \max\!\left(0,\,
  m - f_{\mathrm{rel}}(q,t^+)
    + f_{\mathrm{rel}}(q,t^-)
  \right).
  \label{eq:margin}
\end{equation}
Here $t^+\in\mathcal{R}_q$ is a ground-truth relevant tool. Negatives are
sampled from the first-stage top-100 candidate set whenever possible,
excluding ground-truth tools; if too few negatives are available, we fall
back to corpus-level negatives. This keeps the training objective aligned
with the reranking setting. For risk supervision, we normalize the ordinal
label as $\bar r_i=(r_i-1)/4$ and use mean squared error:
\begin{equation}
  \mathcal{L}_{\mathrm{risk}}
  =
  \frac{1}{|\mathcal{B}_t|}
  \sum_{t_i\in\mathcal{B}_t}
  \left(f_{\mathrm{risk}}(t_i)-\bar r_i\right)^2 ,
  \label{eq:risk_loss}
\end{equation}
where $\mathcal{B}_t$ is the set of tools appearing in the training batch.
We use MSE because the labels are ordinal rather than nominal.

\textbf{Inference-time tradeoff.}
At inference time, the two scores are combined as:
\begin{equation}
  s(q,t)
  =
  f_{\mathrm{rel}}(q,t)
  -
  \lambda f_{\mathrm{risk}}(t),
  \qquad \lambda \ge 0 .
  \label{eq:score_method}
\end{equation}
The parameter $\lambda$ is selected on the validation split and determines
the desired operating point. Setting $\lambda=0$ recovers a risk-blind
reranker; increasing $\lambda$ gives more weight to the learned risk estimate.
Because $\lambda$ is applied only at inference time, the same trained heads
can be used for different relevance--safety operating points.

\subsection{ToolGraph Score Smoothing}
\label{sec:toolgraph}

The dual-head reranker scores each candidate independently. To share ranking
evidence among related tools, we construct an offline ToolGraph
$G=(\mathcal{T},E)$ over the tool library. The graph contains four edge types:
\emph{co-occurrence} edges for tools appearing together in training queries,
\emph{semantic} edges for tools with similar descriptions, \emph{permission}
edges for tools sharing high-risk permission categories, and
\emph{risk-co-occurrence} edges for higher-risk tools that co-occur in
training queries. The graph shares query-specific ranking evidence among related tools and serves as a relational smoothing module rather than a safety constraint; the main safety control comes from the learned risk penalty and the optional rule filter.

For each pair of tools $(t_i,t_j)$, the raw edge weight is the sum of four
type-specific terms:
\begin{equation}
w^{\mathrm{raw}}_{ij}
=
w^{\mathrm{co}}_{ij}
+
w^{\mathrm{sem}}_{ij}
+
w^{\mathrm{perm}}_{ij}
+
w^{\mathrm{risk}}_{ij}.
\label{eq:raw_graph_weight}
\end{equation}
An edge is retained if at least one term is non-zero. We normalize the
retained weights within each dataset:
\begin{equation}
w_{ij}
=
\frac{w^{\mathrm{raw}}_{ij}}
{\max_{(u,v)\in E} w^{\mathrm{raw}}_{uv}} .
\label{eq:graph_weight_norm}
\end{equation}
Co-occurrence and risk-co-occurrence edges are built from training queries
only. Semantic edges are computed from ToolRet-BGE description embeddings.
Permission edges use keyword-derived categories such as shell execution,
file write, network access, credential handling, and code execution. Exact
edge definitions and graph statistics are given in Appendix~\ref{app:graph}.

For a query $q$, let $\mathbf{s}_q$ be the raw score vector from Eq.~\eqref{eq:score_method} over the scored tool set. Propagation is restricted to the subgraph induced by this set, with the corresponding normalized adjacency matrix $\hat{A}_q$; in the main full-pool setting this is the graph over $\mathcal{T}$, and in the candidate-matched setting (Appendix~\ref{app:top100}) it is the top-100-induced subgraph. One-hop
score smoothing is then:
\begin{equation}
  \mathbf{h}_0 = \mathbf{s}_q,\qquad
  \mathbf{h}_1 =
  (1-\alpha)\mathbf{h}_0
  +
  \alpha\,\mathbf{h}_0\hat A_q^{\top},
  \qquad
  \mathbf{s}'_q = \frac{1}{2}(\mathbf{h}_0+\mathbf{h}_1).
  \label{eq:prop}
\end{equation}
The smoothing coefficient $\alpha$ is selected on a held-out validation split
($\alpha=0.2$ for UltraTool and $\alpha=0.02$ for Seal-Tools). We use a
single propagation step to avoid over-smoothing and to keep inference
lightweight.

\subsection{Optional Rule Filter}
\label{sec:rulefilter}

For deployments that require a more conservative exposure policy, we
apply an optional rule filter after reranking. The filter assumes an audited tool registry with
risk-level metadata and permission categories. It scans the reranked candidate
list in order and accepts the first tools that satisfy three constraints.

First, the final top-$K$ list contains at most one higher-risk tool
($r\ge3$). Second, it contains at most two tools matching two or more
high-risk permission categories. Third, a candidate is rejected if its
normalized ToolRet-BGE embedding has cosine similarity greater than $0.9$
with any already selected tool. Accepted tools keep their reranked order. If
fewer than $K$ tools satisfy all constraints, deferred candidates are appended
in their original reranked order until the list reaches length $K$. Thus all
methods are evaluated with the same top-$K$ length.

Under this deployment policy, the risk and permission caps limit higher-risk exposure and privilege breadth, while the similarity threshold controls redundancy. Their values
can be adjusted to deployment needs. The filter adds no trainable parameters. With fixed $K$, the greedy pass
scans each candidate once and compares it with at most $K$ accepted tools, so
the per-query cost is linear in the candidate-list length. Appendix~\ref{app:rule_filter}
gives the exact constraints and pseudocode.

\section{Experiments}
\label{sec:exp}

We conduct experiments to answer the following Research Questions (\textbf{RQs}).

\begin{itemize}[leftmargin=1.5em,itemsep=2pt,topsep=3pt]
  \item \textbf{RQ1:} Does risk-aware reranking improve the relevance--safety
  trade-off compared with relevance-only retrievers and general-purpose
  rerankers?

  \item \textbf{RQ2:} Which components of the framework contribute to
  relevance preservation and risky-tool exposure reduction?

  \item \textbf{RQ3:} How do the tradeoff parameter $\lambda$, ToolGraph
  smoothing, and the rule filter affect different safety--utility operating
  points?

  \item \textbf{RQ4:} What is the tradeoff between reducing unnecessary
risky-tool exposure and retaining genuinely needed high-risk tools?

  \item \textbf{RQ5:} How does the method behave in candidate-exposure and
  scenario-aware stress-test settings?
\end{itemize}

\subsection{Experimental Setup}
\label{sec:setup}

\textbf{Datasets and candidate generation.}
We evaluate on two tool-retrieval benchmarks. UltraTool~\citep{huang2024planning}
contains 2,032 tools across 14 domains with 1,000 test queries and serves as
the primary benchmark. Seal-Tools~\citep{wu2024seal} contains 4,076 API-style
tools. We use the \texttt{test\_in} split with 700 queries for the main comparison and
the \texttt{test\_out} split with 654 queries for out-of-distribution analysis. The eight
general-purpose rerankers use the ToolRet-BGE
top-100 results, while our end-to-end configuration scores the
complete tool pool.

\textbf{Baselines.}
We compare against the retrieval-only ToolRet-BGE baseline and eight
general-purpose rerankers. The reranking baselines include CE-MiniLM-L6,
CE-MiniLM-L12, BGE-Reranker-v2-m3~\citep{xiao2024c}, mxbai-rerank-large
\citep{emb2024mxbai}, MonoT5-base, MonoT5-large~\citep{nogueira2020document},
and Qwen2-0.5B/1.5B-based rerankers~\citep{hui2024qwen2}. The eight reranking baselines receive
the same ToolRet-BGE top-100 candidates.

\textbf{Metrics.}
We report relevance and safety metrics at $k=5$. NDCG@5 and MRR measure
standard ranking quality. To measure retrieval-time risk exposure, we use
RVR@5 and SRR@5. Let $\sigma_q^{(k)}$ be the top-$k$ list returned for query
$q$. Risky-tool Violation Rate is defined as:
\begin{equation}
  \mathrm{RVR@}k =
  \frac{1}{|\mathcal{Q}|}\sum_{q\in\mathcal{Q}}
  \frac{|\{t\in\sigma_q^{(k)}: r_t\ge 3\}|}{k}.
\end{equation}
SRR@5 is the same metric with a stricter threshold $r_t\ge 4$. Lower RVR@5
and SRR@5 indicate less exposure to higher-risk executable tools. We also
report safety-adjusted NDCG, where the relevance gain of severe-risk tools
is removed:
\begin{equation}
  \tilde{g}(t)=g(t)\cdot\mathbf{1}[r_t<4].
\end{equation}
RVR@5 and SRR@5 are the primary safety metrics, while NDCG@5, MRR, and
sNDCG@5 capture relevance and safety-adjusted relevance. The \sndcg{} assigns
zero gain to every L4--L5 tool, but it does not distinguish unnecessary
exposure from tasks that genuinely require such tools. We report Safe-RVR@5 and NeedRisk-Hit@5 in Table~5 to show the tradeoff separately.

\textbf{Implementation details.}
The backbone encoder is a frozen ToolRet-BGE-large model with 1024-dimensional
embeddings. The relevance head takes the concatenated query--tool embedding
as input, while the risk head takes only the tool embedding. Both heads are
two-layer MLPs with hidden size 64 and sigmoid outputs, resulting in 196,866
trainable parameters. We train with Adam using a learning rate of $10^{-3}$ and a batch
size 64, 10 epochs, five negatives per query, margin $m=0.1$, and risk-loss
weight $\mu=0.5$.

\textbf{Risk annotation and evaluation protocol.}
We assign five-level operational risk labels to all 6,108 tools before
training and evaluation. The labels are used as supervision for the risk head,
metadata for graph construction, and evaluation labels for exposure metrics.
Unless otherwise stated, trained variants are evaluated over three random
seeds. Mean values are reported in the main tables for readability.

\subsection{Main Performance Comparison}
\label{sec:main_results}

\begin{table*}[t]
  \caption{
    Main results on UltraTool and Seal-Tools test\_in.
    The eight general-purpose rerankers use ToolRet-BGE top-100 candidates; our methods score the complete tool pool.
    Bold indicates the best result and underlining indicates the second-best result.
    Relative changes for our methods are computed against the strongest non-Ours baseline.
  }
  \label{tab:main}
  \centering
  \begingroup
  \renewcommand{\arraystretch}{1.25}
  \setlength{\tabcolsep}{4.0pt}
  \newcommand{\chg}[2]{\shortstack{#1\\[-1pt]{\scriptsize (#2)}}}
  \resizebox{0.99\textwidth}{!}{%
  \begin{tabular}{lc ccccc ccccc}
  \toprule
  \multirow{2}{*}{Method} & \multirow{2}{*}{Params}
  & \multicolumn{5}{c}{\textsc{UltraTool} (2,032 tools)}
  & \multicolumn{5}{c}{\textsc{Seal-Tools test\_in} (4,076 tools)} \\
  \cmidrule(lr){3-7} \cmidrule(lr){8-12}
  & & NDCG@5$\uparrow$ & MRR$\uparrow$ & sNDCG@5$\uparrow$ & RVR@5$\downarrow$ & SRR@5$\downarrow$
  & NDCG@5$\uparrow$ & MRR$\uparrow$ & sNDCG@5$\uparrow$ & RVR@5$\downarrow$ & SRR@5$\downarrow$ \\
  \midrule

  ToolRet-BGE (no rerank) & 335M
  & 0.429 & 0.499 & 0.426 & 0.176 & 0.078
  & 0.787 & \textbf{0.946} & 0.774 & 0.020 & 0.006 \\

  CE-MiniLM-L6 & 22M
  & 0.334 & 0.427 & 0.322 & 0.178 & 0.089
  & 0.607 & 0.908 & 0.605 & 0.017 & 0.004 \\

  CE-MiniLM-L12 & 33M
  & 0.320 & 0.415 & 0.309 & 0.177 & 0.087
  & 0.527 & 0.819 & 0.525 & 0.016 & 0.003 \\

  MonoT5-base & 220M
  & 0.331 & 0.424 & 0.319 & 0.179 & 0.081
  & 0.775 & 0.912 & 0.784 & 0.018 & 0.003 \\

  MonoT5-large & 770M
  & 0.369 & 0.466 & 0.359 & 0.159 & 0.071
  & 0.767 & 0.924 & 0.755 & 0.018 & 0.003 \\

  BGE-Reranker-v2-m3 & 568M
  & 0.219 & 0.318 & 0.222 & 0.158 & 0.062
  & 0.787 & 0.905 & 0.785 & 0.018 & 0.004 \\

  mxbai-rerank-large & 435M
  & 0.285 & 0.383 & 0.273 & 0.152 & 0.073
  & 0.458 & 0.700 & 0.457 & 0.019 & 0.003 \\

  Qwen2-0.5B-Reranker & 494M
  & 0.275 & 0.339 & 0.261 & \underline{0.119} & \underline{0.042}
  & 0.554 & 0.728 & 0.553 & 0.032 & 0.003 \\

  Qwen2-1.5B-Reranker & 1.5B
  & 0.505 & 0.580 & 0.502 & 0.178 & 0.082
  & 0.646 & 0.820 & 0.644 & 0.026 & 0.003 \\

  \midrule

  Ours (Core + Graph, $\lambda{=}0.1$) & 0.2M
  & \chg{\textbf{0.562}}{$\uparrow$11.3\%}
  & \chg{\textbf{0.625}}{$\uparrow$7.8\%}
  & \chg{\textbf{0.565}}{$\uparrow$12.5\%}
  & \chg{0.145}{$\uparrow$21.8\%}
  & \chg{0.062}{$\uparrow$47.6\%}
  & \chg{\textbf{0.831}}{$\uparrow$5.6\%}
  & \chg{\textbf{0.946}}{$\uparrow$0.0\%}
  & \chg{\textbf{0.829}}{$\uparrow$5.6\%}
  & \chg{\underline{0.011}}{$\downarrow$31.3\%}
  & \chg{\underline{0.003}}{$\uparrow$0.0\%} \\

  Ours + Rule Filter & 0.2M
  & \chg{\underline{0.522}}{$\uparrow$3.4\%}
  & \chg{\underline{0.612}}{$\uparrow$5.5\%}
  & \chg{\underline{0.535}}{$\uparrow$6.6\%}
  & \chg{\textbf{0.073}}{$\downarrow$38.7\%}
  & \chg{\textbf{0.019}}{$\downarrow$54.8\%}
  & \chg{\underline{0.828}}{$\uparrow$5.2\%}
  & \chg{\underline{0.943}}{$\downarrow$0.3\%}
  & \chg{\underline{0.826}}{$\uparrow$5.2\%}
  & \chg{\textbf{0.007}}{$\downarrow$56.3\%}
  & \chg{\textbf{0.002}}{$\downarrow$33.3\%} \\

  \bottomrule
  \end{tabular}%
  }
  \endgroup
\end{table*}

The main comparison shows a consistent relevance--safety trade-off across
the two benchmarks. The core reranker provides the strongest relevance-oriented
operating point, while the rule-filtered variant gives a more conservative
safety-oriented point. Table~\ref{tab:main} compares the eight general-purpose rerankers, which use
ToolRet-BGE top-100 candidates, against our full-pool configuration, which evaluates end-to-end performance without relying on upstream candidate truncation.
Under the same top-100 candidates, Core+Graph obtains 0.621 NDCG@5 (Appendix~\ref{app:top100}).

On UltraTool, the core reranker achieves the best relevance results, with
NDCG@5 of 0.562 and MRR of 0.625. Compared with the strongest
baseline, Qwen2-1.5B-Reranker, it improves ranking quality while training only two
lightweight heads. The rule-filtered variant sacrifices some relevance
but substantially reduces exposure, lowering RVR@5 to 0.073 and SRR@5 to
0.019.

Seal-Tools shows the same division of operating points, although absolute
risk values are smaller because the benchmark contains fewer medium-or-higher
risk tools. The core reranker improves NDCG@5 over ToolRet-BGE, while the
rule-filtered variant further reduces RVR@5. These results suggest that the
core model is better suited for relevance-oriented retrieval, whereas the
rule-filtered model is more appropriate for safety-critical deployments.

\subsection{Component Analysis}
\label{sec:component_analysis}

\begin{table}[t]
  \caption{
  Ablation on UltraTool ($Q{=}1000$, $k{=}5$).
  Each row adds one component over the previous row.
  Rows 2--6 are reported as means over 3 random seeds.
  }
  \label{tab:ablation}
  \centering
  \begingroup
  \renewcommand{\arraystretch}{1.10}
  \setlength{\tabcolsep}{3.5pt}
  \resizebox{\columnwidth}{!}{%
  \begin{tabular}{lccccc}
  \toprule
  Method & NDCG@5$\uparrow$ & sNDCG@5$\uparrow$ & MRR$\uparrow$ & RVR@5$\downarrow$ & SRR@5$\downarrow$ \\
  \midrule

  ToolRet-BGE
  & 0.429 & 0.426 & 0.499 & 0.176 & 0.078 \\

  Risk-label penalty
  & 0.391 & 0.400 & 0.466 & \underline{0.127} & \underline{0.047} \\

  Trained $f_{\mathrm{rel}}$
  & \underline{0.558} & 0.548 & 0.614 & 0.188 & 0.097 \\

  + Trained $f_{\mathrm{risk}}$
  & 0.551 & \underline{0.552} & \underline{0.615} & 0.138 & 0.063 \\

  + Graph smoothing
  & \textbf{0.562} & \textbf{0.565} & \textbf{0.625} & 0.145 & 0.062 \\

  + Rule filter
  & 0.522 & 0.535 & 0.612 & \textbf{0.073} & \textbf{0.019} \\

  \bottomrule
  \end{tabular}%
  }
  \endgroup
\end{table}

\begin{table}[t]
  \caption{
  Edge-type ablation on UltraTool at $\lambda{=}0.1$ and $\alpha{=}0.2$
  (3-seed mean).
  }
  \label{tab:edge_ablation}
  \centering
  \begingroup
  \renewcommand{\arraystretch}{1.10}
  \setlength{\tabcolsep}{8pt}
  \begin{tabular}{lcc}
  \toprule
  Graph configuration & NDCG@5$\uparrow$ & RVR@5$\downarrow$ \\
  \midrule
  No graph ($\alpha{=}0$) & 0.5508 & \textbf{0.1384} \\
  Co-occurrence + semantic & \textbf{0.5703} & 0.1489 \\
  Risk-related only & 0.5508 & 0.1655 \\
  Full 4-type graph & \underline{0.5624} & \underline{0.1449} \\
  \bottomrule
  \end{tabular}
  \endgroup
\end{table}

The ablation study reveals a clear division of labor among the components.
The relevance head is responsible for most of the ranking gain, the risk head
reduces unnecessary exposure, and the rule filter provides the largest exposure reduction. Table~\ref{tab:ablation} quantifies this progression by adding
one component at a time on UltraTool.

Training only the relevance head raises NDCG@5 from 0.429 to 0.558, but it
also increases RVR@5 from 0.176 to 0.188. This failure mode is important:
better semantic matching can promote high-risk tools when they are plausible
matches for the query. Adding the risk head reduces RVR@5 to 0.138 while
maintaining high relevance, showing that risk supervision corrects part of
this exposure without relying on hard rules. Graph smoothing further improves
ranking quality, while the rule filter gives the largest reduction in RVR@5
and SRR@5.

The edge ablation in Table~\ref{tab:edge_ablation} clarifies the role of the
graph, which primarily supports relevance rather than
safety. Relative to no graph propagation, the full graph changes
NDCG@5 from 0.5508 to 0.5624 and RVR@5 from 0.1384 to 0.1449. Co-occurrence and semantic edges provide the strongest relevance gain,
whereas risk-related edges alone do not improve ranking quality. We therefore
treat ToolGraph as a relational smoothing module rather than the primary
safety mechanism; most exposure reduction comes from the learned risk penalty
and the optional rule filter.

\subsection{Relevance--Safety Operating Points}
\label{sec:tradeoff}

\begin{figure*}[t]
  \centering
  \includegraphics[width=\textwidth]{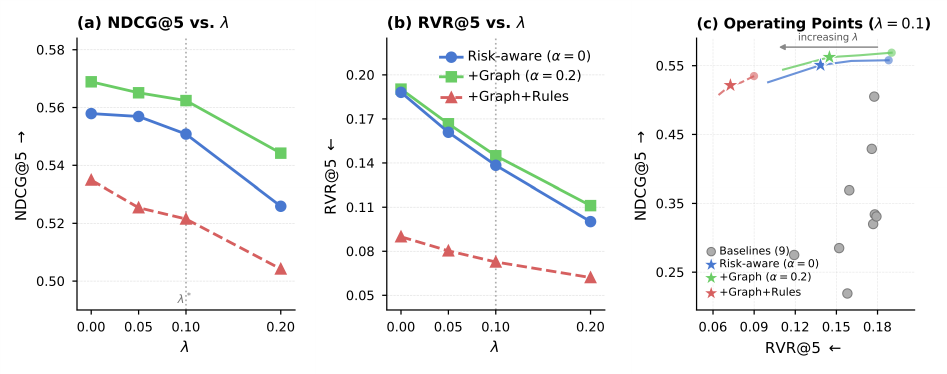}
  \caption{
  Effect of the tradeoff parameter $\lambda$ on UltraTool.
  (a)--(b): NDCG@5 and RVR@5 as a function of $\lambda$ for three
  configurations. Shaded regions indicate $\pm$1 std over 3 seeds.
  The dashed vertical line marks $\lambda^{*}=0.1$.
  (c): Baselines and our configurations on the NDCG@5--RVR@5 plane.
  Arrows indicate increasing $\lambda$.
  }
  \Description{Line plots showing the effect of lambda on NDCG@5 and RVR@5,
  and a scatter plot of operating points on the NDCG@5--RVR@5 plane.}
  \label{fig:tradeoff}
\end{figure*}

The tradeoff parameter $\lambda$ gives a continuous way to move between
relevance-oriented and safety-oriented operating points. We vary
$\lambda \in \{0,0.05,0.1,0.2\}$ for the risk-aware, graph-smoothed, and
rule-filtered configurations; the resulting curves are shown in
Figure~\ref{fig:tradeoff}.

Increasing $\lambda$ consistently reduces risky-tool exposure. In the
risk-aware and graph-smoothed configurations, RVR@5 decreases from 0.188 to
0.100 and from 0.190 to 0.111, respectively, as $\lambda$ increases from 0 to
0.2. This reduction comes with a moderate decrease in NDCG@5, from 0.558 to
0.526 and from 0.569 to 0.544. The rule-filtered configuration stays at a
lower-risk operating point across the full range of $\lambda$, because the
rule filter is applied on top of the learned risk penalty.

Figure~\ref{fig:tradeoff}(c) shows that the
baselines cluster at higher RVR@5 values, while our variants move toward lower
exposure as $\lambda$ increases. The parameter $\lambda$ and the rule filter
therefore play complementary roles: $\lambda$ provides a smooth tuning knob,
whereas the rule filter defines a conservative deployment setting.

\subsection{Context-Aware Safety and Generalization}
\label{sec:robustness}

\begin{figure*}[t]
  \centering
  \includegraphics[width=\textwidth]{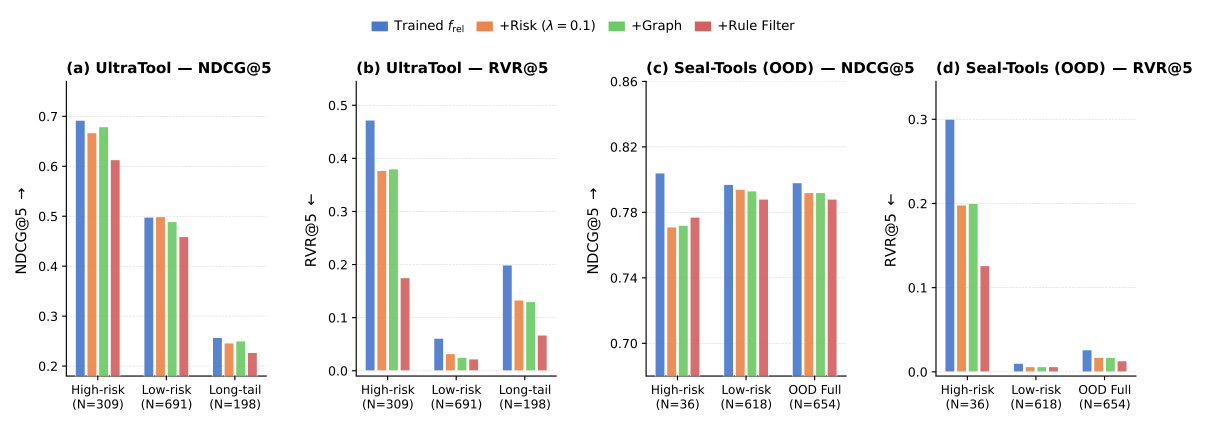}
  \caption{
  NDCG@5 and RVR@5 across query subsets on UltraTool and Seal-Tools.
  High-/low-risk queries are split by the maximum risk level of ground-truth
  tools; long-tail queries target tools appearing in at most 5 training queries.
  }
  \Description{Bar charts comparing NDCG@5 and RVR@5 across high-risk,
  low-risk, long-tail, and out-of-distribution query subsets.}
  \label{fig:robustness}
\end{figure*}

\begin{table}[t]
  \caption{
  Context-aware safety metrics. Safe-RVR@5 is computed on low-risk queries;
  NeedRisk-Hit@5 is computed on high-risk queries.
  }
  \label{tab:context_safety}
  \centering
  \begingroup
  \renewcommand{\arraystretch}{1.10}
  \setlength{\tabcolsep}{6pt}
  \begin{tabular}{llcc}
  \toprule
  Dataset & Method & Safe-RVR@5$\downarrow$ & NeedRisk-Hit@5$\uparrow$ \\
  \midrule

  \multirow{4}{*}{UltraTool}
  & Trained $f_{\mathrm{rel}}$ & 0.0260 & \textbf{0.660} \\
  & +Risk  & \underline{0.0147} & \underline{0.506} \\
  & +Graph & \underline{0.0147} & 0.489 \\
  & +Rules & \textbf{0.0102} & 0.397 \\

  \midrule

  \multirow{4}{*}{Seal-Tools}
  & Trained $f_{\mathrm{rel}}$ & 0.0094 & \textbf{0.727} \\
  & +Risk  & \underline{0.0029} & \underline{0.576} \\
  & +Graph & \underline{0.0029} & \underline{0.576} \\
  & +Rules & \textbf{0.0027} & 0.545 \\

  \bottomrule
  \end{tabular}
  \endgroup
\end{table}

A useful safety mechanism should not simply suppress every high-risk tool:
some queries genuinely require higher-risk operations. We therefore split
queries according to whether their ground-truth tool set contains a tool with
$r\geq3$, and separately evaluate long-tail and out-of-distribution queries.

The subset results in Figure~\ref{fig:robustness} show that most exposure
reduction occurs where risk is actually present. On UltraTool high-risk
queries, adding the risk head reduces RVR@5 from 0.472 to 0.377, and the rule
filter lowers it further to 0.175. Low-risk queries show smaller changes in
NDCG@5, suggesting that the method mainly removes unnecessary risky tools
rather than uniformly degrading retrieval.

To further evaluate the safety and robustness of tool retrieval, we propose two additional metrics, Safe-RVR@5 and NeedRisk-Hit@5. Let $Q_{\mathrm{safe}}=\{q:\max_{t\in R_q} r_t < 3\}$ and
$Q_{\mathrm{need}}=\{q:\exists t\in R_q,\ r_t\ge 3\}$. We define:
\begin{equation}
\mathrm{Safe\text{-}RVR}@k =
\frac{1}{|Q_{\mathrm{safe}}|}
\sum_{q\in Q_{\mathrm{safe}}}
\frac{|\{t\in \sigma_q^{(k)}: r_t\ge 3\}|}{k},
\end{equation}
and
\begin{equation}
\begin{aligned}
\mathrm{NeedRisk\text{-}Hit}@k
= \frac{1}{|\mathcal{Q}_{\mathrm{need}}|}
\sum_{q\in\mathcal{Q}_{\mathrm{need}}}
\mathbf{1}\!\big[
&\exists\, t \in \sigma_q^{(k)} \cap \mathcal{R}_q \\
&\mathrm{s.t.}\; r_t \ge 3
\big].
\end{aligned}
\end{equation}
These two metrics can measure the trade-off between context-aware safety and the generalization capability. As shown in Table~\ref{tab:context_safety}, as safety controls become stricter, Safe-RVR@5 decreases,
which means fewer high-risk tools are exposed on low-risk queries. At the
same time, NeedRisk-Hit@5 also decreases, especially under the rule filter.
This is the expected cost of conservative filtering: it can remove high-risk
tools even when they are genuinely needed. We therefore treat $\lambda$ and
the rule filter as deployment controls rather than universally optimal
settings.

The Seal-Tools \texttt{test\_out} results provide a separate check on
generalization. Since these tools are unseen during training, the reduction in
RVR@5 indicates that the learned risk head uses tool descriptions rather than
memorizing tool identities.

\subsection{Candidate Exposure and Stress Tests}
\label{sec:case_study}

\begin{table}[t]
  \caption{
  Candidate exposure analysis without execution on UltraTool.
  The agent sees only the query and the top-5 candidate tools.
  RVR@5 and SRR@5 measure the risk exposure of the candidate set.
  }
  \label{tab:tool_selection}
  \centering
  \begingroup
  \renewcommand{\arraystretch}{1.10}
  \setlength{\tabcolsep}{3.5pt}
  \resizebox{\columnwidth}{!}{%
  \begin{tabular}{lccccc}
  \toprule
  Method & NDCG@5$\uparrow$ & sNDCG@5$\uparrow$ & MRR$\uparrow$ & RVR@5$\downarrow$ & SRR@5$\downarrow$ \\
  \midrule

  BM25
  & 0.2876 & 0.2829 & 0.3643 & 0.1962 & 0.0990 \\

  ToolRet-BGE
  & \underline{0.4288} & \underline{0.4258} & \underline{0.4993} & 0.1758 & 0.0784 \\

  ToolRet-BGE + MonoT5-large
  & 0.3690 & 0.3594 & 0.4657 & 0.1594 & 0.0710 \\

  ToolRet-BGE + Rule Filter
  & 0.4103 & 0.4181 & 0.4895 & \textbf{0.0662} & \underline{0.0228} \\

  Ours + Graph + Rule Filter
  & \textbf{0.5215} & \textbf{0.5351} & \textbf{0.6120} & \underline{0.0727} & \textbf{0.0191} \\

  Ours ($\lambda{=}0.10$)
  & -- & -- & -- & 0.1357 & 0.0395 \\

  \bottomrule
  \end{tabular}%
  }
  \endgroup
\end{table}

\begin{table}[t]
  \caption{
    Stress-test results on UltraTool ($n{=}30$).
    Results are reported as means over 3 seeds.
    }
  \label{tab:stress}
  \centering
  \begingroup
  \renewcommand{\arraystretch}{1.10}
  \setlength{\tabcolsep}{8pt}
  \begin{tabular}{lccc}
  \toprule
  Method & NDCG@5$\uparrow$ & RVR@5$\downarrow$ & SRR@5$\downarrow$ \\
  \midrule

  ToolRet-BGE
  & \textbf{0.658} & 0.393 & 0.233 \\

  +Graph Smoothing
  & \underline{0.562} & \underline{0.324} & \underline{0.160} \\

  +Rule Filter
  & 0.478 & \textbf{0.111} & \textbf{0.042} \\

  \bottomrule
  \end{tabular}
  \endgroup
\end{table}

The previous experiments evaluate standard retrieval benchmarks. We next
isolate the candidate action space that would be shown to an agent before any
tool is executed. This diagnostic setting does not evaluate tool-call outcomes;
instead, it asks whether different retrieval and reranking strategies expose
the agent to different levels of tool risk.

Table~\ref{tab:tool_selection} compares sparse retrieval, dense retrieval,
general-purpose reranking, rule-based filtering, and our risk-aware pipeline
on UltraTool. Relevance-driven retrieval exposes a non-trivial number of
higher-risk tools: ToolRet-BGE obtains RVR@5 of 0.1758 and SRR@5 of 0.0784.
MonoT5-large slightly reduces exposure, but does not directly optimize the
risk profile of the candidate set. Rule filtering yields the lowest RVR@5,
while our graph-enhanced rule-filtered variant achieves the strongest
relevance metrics and the lowest SRR@5. Without the rule filter, the
learned risk-aware reranker also reduces exposure relative to ToolRet-BGE,
lowering RVR@5 from 0.1758 to 0.1357 and SRR@5 from 0.0784 to 0.0395.

We further construct stress-test queries in which the task-relevant tool is
safe or appropriate, but semantically similar risky tools are retrieved by the
first-stage retriever. On UltraTool, the rule-filtered system reduces RVR@5
from 0.393 to 0.111 and SRR@5 from 0.233 to 0.042. This stress setting
highlights the intended role of retrieval-stage control: reducing the density
of risky candidates before the agent chooses or executes a tool.




\subsection{Qualitative Downstream Inspection}

We further conduct a qualitative downstream inspection to examine whether the
rule-filtered top-5 lists align with the intended pre-execution safety
objective. Rather than listing every retrieved candidate,
Table~\ref{tab:case_study} reports two signals for six representative queries:
the number of gold tools retained after filtering and the risk levels of the
five selected candidates in rank order. This inspection serves as a sanity
check on whether the filter reduces unnecessary risky-tool exposure while
preserving task-relevant tools.

The first three cases are high-consequence financial workflows. In all three,
the filtered list contains only one L3 tool and no L4/L5 tools. For example,
in the credit-card repayment case, the filter preserves the necessary login
and debt-query tools while avoiding additional high-risk financial operations
that are not needed in the top-5 candidate set. This suggests that the filter
can reduce risk density when multiple semantically related financial tools are
available but only a small subset is required.

The last three cases involve routine service or file-operation workflows. Here,
the filter may still retain one L4 tool, such as \texttt{file\_delete}, when it
is directly relevant to the user request. Thus, the rule filter does not impose
a rigid zero-risk policy. Instead, it suppresses redundant or unnecessary
high-risk tools while allowing task-critical tools to remain. Overall, these
cases support the rule-filtered variant as a conservative pre-execution control
rather than a blanket safety blocker.

\begin{table}[t]
\caption{
Qualitative downstream candidate inspection after rule filtering.
Risk profiles list the five filtered candidates' risk levels in rank order.
}
\label{tab:case_study}
\centering
\small
\setlength{\tabcolsep}{3.5pt}
\renewcommand{\arraystretch}{1.08}
\resizebox{\columnwidth}{!}{%
\begin{tabular}{llccc}
\toprule
Setting & Scenario & Gold retained & Risk profile & Takeaway \\
\midrule
Finance & Investment submission & 1/4 & [3,2,2,1,2] & No L4/L5 \\
Finance & Credit-card repayment & 2/3 & [3,2,2,1,2] & No L4/L5 \\
Finance & Foreign-currency purchase & 2/3 & [3,2,1,2,1] & No L4/L5 \\
Service & Hotel booking modification & 1/2 & [1,2,4,1,2] & Keeps one L4 \\
File operation & Price-list edit/delete & 1/2 & [4,2,2,1,2] & Keeps gold L4 \\
File operation & Modify then delete file & 1/2 & [4,2,2,1,2] & Keeps gold L4 \\
\bottomrule
\end{tabular}}
\end{table}

\section{Conclusion}
\label{sec:conclusion}

We present a lightweight risk-aware reranking framework for agent tool
retrieval. The framework treats retrieval as a pre-execution safety boundary:
before an LLM agent invokes any tool, the retrieved top-$k$ list already
defines the candidate action space exposed to the agent. By separating query-conditioned relevance from tool-level exposure risk, our method provides a controllable safety--utility tradeoff on top of a frozen first-stage retriever. Experiments on UltraTool and Seal-Tools show that the
proposed reranker reduces risky-tool exposure while maintaining competitive
retrieval quality, and that the rule-filtered variant offers a conservative deployment setting.

We acknowledge several limitations. The risk model estimates tool-level
operational risk from tool descriptions and does not capture all
query-level, user-specific, or environment-dependent hazards that may arise
during actual execution. Learning
$f_{\mathrm{risk}}(q,t)$ would require per-$(q,t)$ risk supervision,
whereas our released annotations are defined per tool. In addition, the rule-filtered mode assumes an
audited tool registry with available risk and permission metadata. Therefore,
risk-aware retrieval should be viewed as a complementary upstream safeguard
rather than a replacement for downstream permission control, user confirmation,
and execution-time safety checks.


\appendix

\section{Risk Annotation Details}
\label{app:risk_annotation}

This appendix summarizes the annotation protocol used to produce the
five-level tool-risk labels.

\paragraph{Annotation protocol.}
All the annotators label each tool from its name and description. They consider three
axes: the scope of impact, the reversibility of the action, and the sensitivity
of the accessed or modified information. The prompt asks LLM annotators to assign
a single integer from 1 to 5 according to the following rubric: L1 for
read-only tools with no side effects, L2 for minor reversible actions or
non-sensitive personal-data access, L3 for sensitive data access or persistent
writes, L4 for irreversible actions or system-level permissions, and L5 for
large-scale harm, system intrusion, or severe privacy loss.

\paragraph{Prompt template.}
Each LLM annotator is given the same prompt template:
\begin{quote}
\small
Given the following tool name and description, assign one risk level from
L1 (Safe) to L5 (Critical). Consider scope of impact, reversibility, and data
sensitivity. Return only one integer in \{1,2,3,4,5\}.

\textbf{Tool name:} \texttt{\{tool\_name\}}

\textbf{Description:} \texttt{\{tool\_description\}}
\end{quote}

\paragraph{Disagreement resolution.}
Each tool first receives preliminary labels from three independent LLM annotators. Let the three votes be
$v_1,v_2,v_3$, and define
\begin{equation}
\mathrm{span}
=
\max(v_1,v_2,v_3)-\min(v_1,v_2,v_3).
\end{equation}
If $\mathrm{span}=0$, the label is accepted directly. If
$\mathrm{span}=1$, we use the median vote. If $\mathrm{span}\geq2$, the case
is escalated to human review. This rule keeps adjacent-level disagreements
lightweight while manually auditing broad cross-level disagreements.

\begin{table}[t]
\caption{Distribution of resolved tool-risk labels.}
\label{tab:risk_distribution}
\centering
\small
\setlength{\tabcolsep}{6pt}
\renewcommand{\arraystretch}{1.05}
\begin{tabular}{lcc}
\toprule
Risk level & UltraTool & Seal-Tools \\
\midrule
L1 Safe     & 958 (47.1\%)  & 2807 (68.9\%) \\
L2 Low      & 799 (39.3\%)  & 1197 (29.4\%) \\
L3 Medium   & 155 (7.6\%)   & 51 (1.3\%) \\
L4 High     & 115 (5.7\%)   & 20 (0.5\%) \\
L5 Critical & 5 (0.2\%)     & 1 (0.02\%) \\
\midrule
Total       & 2{,}032       & 4{,}076 \\
\bottomrule
\end{tabular}
\end{table}

Seal-Tools contains substantially fewer medium-or-higher-risk tools than
UltraTool, which helps explain its lower absolute RVR@5 and SRR@5 values in
the main experiments.

\section{ToolGraph Construction Details}
\label{app:graph}

This appendix gives the deterministic construction rules for the ToolGraph
used in Section~\ref{sec:toolgraph}. We construct an undirected graph
$G=(\mathcal{T},E)$ over tools. For each pair $(t_i,t_j)$, the raw edge weight
is
\begin{equation}
w^{\mathrm{raw}}_{ij}
=
w^{\mathrm{co}}_{ij}
+
w^{\mathrm{sem}}_{ij}
+
w^{\mathrm{perm}}_{ij}
+
w^{\mathrm{risk}}_{ij},
\end{equation}
and an edge is retained whenever at least one term is non-zero. The retained
weights are normalized by the maximum raw edge weight in the dataset:
\begin{equation}
w_{ij}
=
\frac{w^{\mathrm{raw}}_{ij}}
{\max_{(u,v)\in E} w^{\mathrm{raw}}_{uv}} .
\end{equation}

\begin{table}[t]
\caption{ToolGraph edge definitions.}
\label{tab:graph_edge_definitions}
\centering
\small
\setlength{\tabcolsep}{3pt}
\renewcommand{\arraystretch}{1.05}
\resizebox{\columnwidth}{!}{%
\begin{tabular}{lll}
\toprule
Type & Trigger & Weight \\
\midrule
Co-occurrence
& Co-occur in at least 2 training queries
& $0.4(c_{ij}/c_{\max})$ \\
Semantic
& $\cos(\mathbf{d}_i,\mathbf{d}_j)>0.65$
& $0.3\cos(\mathbf{d}_i,\mathbf{d}_j)$ \\
Permission
& Share high-risk permission category
& $0.15(m_{ij}/5)$ \\
Risk co-occurrence
& Both $r\geq3$ and co-occur in training
& $0.15$ \\
\bottomrule
\end{tabular}}
\end{table}

Here $c_{ij}$ is the number of training queries in which $t_i$ and $t_j$
co-occur, $c_{\max}$ is the maximum co-occurrence count, $\mathbf{d}_i$ is the
ToolRet-BGE description embedding, and $m_{ij}$ is the number of shared
high-risk permission categories.

\paragraph{Permission categories.}
For permission-overlap edges and the rule filter, we use five deterministic
keyword-derived categories: shell execution, file write, network access,
credential handling, and code execution. Example keywords include
\texttt{shell}, \texttt{bash}, \texttt{command}, \texttt{delete},
\texttt{overwrite}, \texttt{upload}, \texttt{download}, \texttt{token},
\texttt{password}, \texttt{oauth}, \texttt{eval}, and \texttt{run\_code}.
These categories are used only as transparent metadata for graph construction
and rule filtering; they are not the learned risk model.

\section{Rule Filter Details}
\label{app:rule_filter}

The rule filter is an optional deployment-time exposure-control layer applied
after risk-aware reranking. It assumes an audited tool registry with risk
metadata and permission categories, and adds no trainable parameters. The fallback was not triggered on either benchmark in our experiments. 
\subsection{Constraints}

Given a reranked candidate list, the filter greedily constructs a top-$K$ list
$S$ subject to three constraints.

\textbf{Risk cap.}
At most one higher-risk tool may appear in the final top-$K$ list:
\begin{equation}
\sum_{t\in S} \mathbf{1}[r_t\geq3] \leq 1 .
\end{equation}

\textbf{Permission cap.}
Let $\mathrm{perm}(t)$ be the number of high-risk permission categories matched
by tool $t$. The filter allows at most two tools with two or more such
categories:
\begin{equation}
\sum_{t\in S} \mathbf{1}[\mathrm{perm}(t)\geq2] \leq 2 .
\end{equation}

\textbf{Redundancy constraint.}
Let $\mathbf{e}_t$ be the normalized ToolRet-BGE embedding of tool $t$. A
candidate is rejected if it is too similar to an already selected tool:
\begin{equation}
\max_{s\in S} \mathbf{e}_t^{\top}\mathbf{e}_s > 0.9 .
\end{equation}

\subsection{Greedy Filtering Procedure}

The filter scans the reranked list once while maintaining an accepted list $S$
and a deferred list $D$:
\begin{enumerate}[leftmargin=1.5em,itemsep=1pt,topsep=2pt]
  \item Initialize $S\leftarrow[\,]$ and $D\leftarrow[\,]$.
  \item For each candidate $t$ in reranked order, append $t$ to $S$ if adding
  it satisfies the risk cap, permission cap, and redundancy constraint.
  Otherwise, append $t$ to $D$.
  \item Stop the acceptance pass once $|S|=K$ or all candidates have been
  scanned.
  \item If $|S|<K$, append candidates from $D$ in their original reranked order
  until $|S|=K$.
\end{enumerate}

The fallback step ensures that all methods produce the same top-$K$ length.
For metrics at fixed $K$, only the returned prefix $S$ is evaluated.

\subsection{Complexity}

The greedy pass scans the candidate list once and compares each candidate with
at most $K$ accepted tools. The per-query complexity is therefore $O(NK)$ for a
candidate list of length $N$. Since $K=5$ in our evaluation, the filter is
linear in the candidate-list length.

\section{Head Architecture and Training Details}
\label{app:head_training}

The ToolRet-BGE encoder is frozen, and only two lightweight MLP heads are
trained. The relevance head maps the concatenated query--tool embedding
through a $2048{\rightarrow}64{\rightarrow}1$ MLP, while the risk head maps
the tool embedding through a $1024{\rightarrow}64{\rightarrow}1$ MLP. Both
heads use sigmoid outputs. The total number of trainable parameters is
196{,}866.

We train the heads with Adam using a learning rate of $10^{-3}$ and a batch size of 64,
10 epochs, five negatives per query, margin $m=0.1$, and risk-loss weight
$\mu=0.5$.

As a sanity check, the mean predicted risk score increases monotonically from
L1 to L5 on UltraTool, indicating that the risk head preserves the ordinal
structure of the labels.

\section{Additional Experimental Results}
\label{app:additional_results}

\subsection{Standard Deviations}

\begin{table}[!htbp]
  \caption{
  Standard deviations of our trained variants over 3 seeds.
  The corresponding mean values are reported in Table~\ref{tab:main}.
  }
  \label{tab:std}
  \centering
  \scriptsize
  \setlength{\tabcolsep}{2.6pt}
  \renewcommand{\arraystretch}{1.05}
  \resizebox{\columnwidth}{!}{%
  \begin{tabular}{llccccc}
  \toprule
  Dataset & Method & NDCG@5 & MRR & sNDCG@5 & RVR@5 & SRR@5 \\
  \midrule
  \multirow{2}{*}{UltraTool}
  & Core + Graph & 0.039 & 0.036 & 0.046 & 0.028 & 0.027 \\
  & + Rule Filter & 0.047 & 0.042 & 0.060 & 0.011 & 0.012 \\
  \midrule
  \multirow{2}{*}{Seal-Tools}
  & Core + Graph & 0.005 & 0.001 & 0.004 & 0.004 & 0.001 \\
  & + Rule Filter & 0.003 & 0.001 & 0.002 & 0.003 & 0.001 \\
  \bottomrule
  \end{tabular}%
  }
\end{table}

\subsection{Candidate-Matched Top-100 Evaluation}\label{app:top100}\par Using the same ToolRet-BGE top-100 candidates as the reranking baselines, Core, Core+Graph, and Core+Graph+Rule obtain NDCG@5 of $0.609\pm0.029$, $0.621\pm0.024$, and $0.573\pm0.032$, respectively. Graph smoothing is normalized on the candidate-induced subgraph. 

\begin{acks}
This work was supported in part by the \grantsponsor{GS501100001809}{National Natural Science Foundation of China}{https://doi.org/10.13039/501100001809} under Grant \grantnum{GS501100001809}{72401282}.
\end{acks}

\clearpage
\section*{GenAI Usage Disclosure}

Claude Code, Codex, and Qwen were used only as preliminary annotators for
tool-risk labels under the fixed rubric in Section~\ref{sec:annotation}.
Labels were reconciled by the span-based procedure, and broad disagreements
were reviewed by human authors. The authors take full responsibility for the
final labels, experiments, and manuscript.

\bibliographystyle{ACM-Reference-Format}
\bibliography{references}

\end{document}